# Strong optomechanical coupling in a slotted photonic crystal nanobeam cavity with an ultrahigh quality factor-to-mode volume ratio

Katharina Schneider[*] and Paul Seidler

*IBM-Research – Zurich, Säumerstrasse 4, CH-8803 Rüschlikon, Switzerland*
[*]*ksc@zurich.ibm.com*

**Abstract:** We describe the design, fabrication, and characterization of a one-dimensional silicon photonic crystal cavity in which a central slot is used to enhance the overlap between highly localized optical and mechanical modes. The optical mode has an extremely small mode volume of 0.017 $(\lambda_{vac}/n)^3$, and an optomechanical vacuum coupling rate of 310 kHz is measured. With optical quality factors up to $1.2 \times 10^5$, fabricated devices are in the resolved-sideband regime. The electric field has its maximum at the slot wall and couples to the in-plane breathing motion of the slot. The optomechanical coupling is thus dominated by the moving-boundary effect, which we simulate to be six times greater than the photoelastic effect, in contrast to most structures, where the photoelastic effect is often the primary coupling mechanism.

## 1. Introduction

Increasingly sophisticated fabrication methods have led in recent years to the ability to make ever-smaller optomechanical systems [1]. Interaction of small mechanical resonators with a strongly confined electromagnetic field enables light modulation at frequencies up to several gigahertz. Such high frequencies are required for applications in communication [2], radio astronomy [3] or the transduction of superconducting qubits [4,5]. Only a small number of platforms have been developed, for which optomechanical modulation occurs at frequencies above 1 GHz. They include one-dimensional (1D) photonic crystals [6-13], two-dimensional photonic crystals [14,15] and disk resonators [16]. Most recently, an approach has been presented, where propagating acoustic waves with frequencies up to 12 GHz exhibit optomechanical coupling with a photonic cavity [17].

In a system consisting of an optical resonator with resonance frequency $\omega_o$ coupled to a mechanical resonator with resonance frequency $\Omega_m$, the magnitude of the effects caused by radiation pressure forces, e.g. the amplitude of the optical sidebands created by thermal motion, scales with the optomechanical vacuum coupling rate $g_0$, which is defined as the shift of $\omega_o$ with a change in the generalized displacement $x$ of the mechanical resonator, namely $g_0 = \frac{\partial \omega_o}{\partial x} x_{zpf}$ in units of the mechanical zero-point fluctuation $x_{zpf}$. The zero-point fluctuation of the mechanical resonator is given by $x_{zpf} = \sqrt{\hbar/2m_{eff}\Omega_m}$, where $m_{eff}$ is the effective mass and $\hbar$ is Planck's constant divided by $2\pi$.

Various physical phenomena can produce optomechanical coupling [1,18,19]. At the nanometer scale, the most commonly considered mechanisms are the moving boundary effect [20] and the photoelastic effect [21]. The former occurs if a dielectric boundary moves through a region of non-zero electric field. The latter is induced by a change of the refractive index with strain. The overall coupling rate is described by the sum of the respective coupling rates for these effects, so that $g_0 = g_{0,mb} + g_{0,pe}$. The relative magnitude of the two effects depends on the optical frequency, with the photoelastic effect being more strongly dispersive, particularly near the electronic bandgap (see Appendix I). Depending on the wavelength of operation, the cavity may be designed to take advantage of one or the other effect. To exploit primarily the moving boundary contribution, slotted photonic crystals have proven to be fruitful [22-25], but most published results involve mechanical frequencies of at most a few tens of megahertz. An exception is the work of Grutter et al. [13] based on separate mechanical and optical beams made of Si$_3$N$_4$.

In addition to the vacuum coupling rate $g_0$, the behavior of an optomechanical system is governed by its optical decay rate $\kappa$, which is related to the optical quality factor $Q_o$ by $Q_o = \omega_o/\kappa$. A small decay rate can be advantageous for certain applications. For example, the ability to achieve ground-state cooling depends on the sideband suppression factor $\kappa/\Omega_m$. Furthermore, for efficient transduction from individual phonons to photons, a high single-photon cooperativity $C = g_0^2/(\kappa \Gamma_m)$ is required, $\Gamma_m$ being the decay rate of the mechanical resonator. For some time, it seemed inherently difficult to achieve a high optical quality factor in cavities comprising a slot because of photon scattering at the slot, and up until now no slotted photonic crystal cavity has achieved the resolved-sideband regime, where $\kappa/\Omega_m > 1$.

In this work we present a system that combines strong optomechanical coupling with a high optical quality factor and an extremely small mode volume. The slotted 1D photonic crystal design possesses an optomechanical coupling rate of $g_0/2\pi = 310$ kHz, a calculated mode volume of $V = 0.017\ (\lambda_{vac}/n)^3$, and a measured optical quality factor above $1.2 \times 10^5$. The mechanical mode investigated is at $\Omega_m/2\pi = 2.69$ GHz, which puts the system in the resolved sideband regime. The device has a small footprint, operates in the telecommunication window with a measured wavelength of $\lambda_{vac} \approx 1544$ nm and can be integrated into on-chip photonic circuits. For instance, by incorporating the photonic crystal cavity into one arm of a Mach-Zehnder interferometer, we are able to utilize homodyne detection and profit from an improved signal-to-noise ratio. Devices are fabricated with varying dimensions to an accuracy of $\lesssim 10$ nm and their mechanical and optical properties characterized. Two independent methods are employed to determine $g_0$: calibration tone and optomechanically induced absorption measurements.

## 2. Working Principle

Displacement of a dielectric interface in a photonic cavity leads to a change in the cavity's optical eigenfrequencies, the so-called moving boundary effect. The magnitude of the frequency shift depends on the spatial distribution of both the optical and mechanical modes. Increased electric field energy in the region of greatest mechanical motion is generally expected to enhance the contribution of the moving boundary effect to the overall optomechanical coupling [22]. To this end, structures incorporating a slot are particularly interesting. Because of the requirement for continuity of the normal component of the electric displacement field across a dielectric boundary, an abrupt change in the normal component of the electric field occurs in slot structures [26] – an effect that has been exploited, for example, to substantially increase the confinement of light in slotted waveguides [27]. For electric fields perpendicular to the slot axis, the maximum is directly at the surface of the slot wall, from where it decays exponentially towards the middle of the slot. Narrow slots are therefore preferable for maintaining a high electrical field throughout the slot.

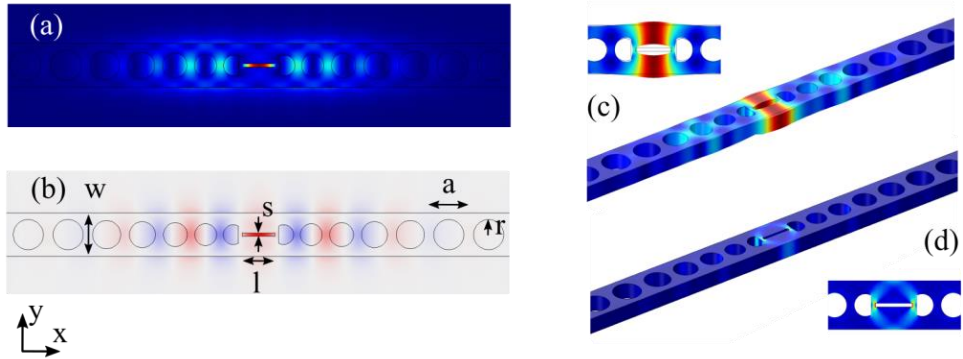

Fig. 1. (a) and (b) Cross-sections through the central portion of the photonic crystal cavity taken from a finite element simulation of the electric field of the highest frequency cavity mode with an antinode in the slot. (a) Total magnitude of the electric field, increasing from blue to red. (b) $E_y$ component of the electric field (red and blue correspond to opposite signs). The simulations indicate that the electric field amplitude for this mode is strongly enhanced in the slot. (c) Displacement image from a finite-element simulation of the in-plane mechanical breathing mode at 2.69 GHz. The displacement overlaps spatially with the region of maximum electric field shown in (a). (d) Distribution of the first principal strain for the breathing mode. The region of greatest strain is nearly identical to the location of the first node in the electric field.

With these considerations in mind, we have chosen to investigate a 1D photonic crystal nanobeam cavity with a central slot (Fig 1). It comprises a freestanding waveguide with a sequence of holes forming Bragg mirrors on either side of the cavity. The unit cell of the periodic holes is scaled linearly in terms of length and hole radius for the five holes closest to the slot so as to ensure a gradual matching of the mode profile between the Bragg mirrors and the slot region, which in turn yields a high optical quality factor [28,29].

Such a 1D photonic crystal nanobeam cavity supports multiple optical modes, the highest frequency mode being symmetric with an antinode of the electric field in the center of the structure. The next mode is antisymmetric with a node of zero electric field in the middle of the device [30]. As the optical cavity modes are confined primarily to the slot region, a mechanical mode overlapping well with the electric field is the in-plane breathing mode involving lateral opening and closing of the slot. In principle, optomechanical coupling can be achieved with both optical modes, but for a large moving boundary contribution, the first mode with an antinode in the middle has greater net overlap with the mechanical breathing mode.

In contrast to our previous work [30], the slot is terminated before the first hole on either side of the cavity in order to increase the mechanical resonance frequency of the breathing mode to several gigahertz. The distance between the first two holes corresponds to roughly half the effective wavelength of the central portion of the structure, so the first nodes in the electric field fall at the crosspieces closing off the first holes. This is also the location of the "joint" for the mechanical breathing mode, the region of greatest deformation. Often the sign of the photoelastic contribution to the optomechanical coupling rate is the opposite of that of the moving boundary contribution, leading to at least a partial cancellation of the effects. In this case, however, the region of most strain is where the electric field is small and the photoelastic and moving boundary contributions have the same sign. The result is a relatively high overall optomechanical coupling rate dominated by the moving boundary contribution, for which the mechanical frequency is high enough to be in the resolved side-band regime.

The final design has a simulated mode volume of $V = 0.017 \ (\lambda_{vac}/n)^3$, which is close to the mode volume published in [30] with a theoretical optical quality factor $Q_o = 1.6 \times 10^6$, which corresponds to $Q_o/V = 9.5 \times 10^7$. The simulated electric field distribution is shown in Fig. 1(a) and (b). The very low effective mass $m_{eff} = 103$ fg leads to a high mechanical resonance frequency of $\Omega_m/2\pi = 2.69$ GHz. We show the simulated displacement in Fig. 1(c) and the associated strain in Fig. 1(d). Analysis of the numerical simulations predicts a vacuum coupling rate of $g_0/2\pi = 342$ kHz. Notably, the moving boundary contribution, $g_{0,mb} = 294$ kHz, is approximately six times larger than the photoelastic contribution, $g_{0,pe} = 47$ kHz.

## 3. Methods

### 3.1 Design

The device design is initially optimized for a high optical quality factor using finite-difference time-domain simulations [31] carried out with the freely available software package MEEP [32]. To optimize the mechanical properties as well, simulations are performed with the multiphysics solver COMSOL [33]. Both the photoelastic effect and the moving boundary effect are considered for the calculation of the overall optomechanical coupling rate $g_0 = g_{0,mb} + g_{0,pe}$ (see Appendix I). A Nelder-Mead algorithm is applied to optimize the geometry of the structure for the fitness function $F = g_0 \min\{Q_o, Q_{lim}\}$ [6]. $Q_{lim} = 1.5 \times 10^6$ is an upper limit for the optical quality factor that prevents the optimization algorithm from considering unreasonably high values calculated by COMSOL. The design produced by the optimization algorithm is scaled so as to provide an optical resonance frequency of approximately $\omega_o/2\pi = 193.5$ THz, which corresponds to $\lambda_{vac} = 1550$ nm. The resulting design has the following dimensions: The holes have a nominal period of $a = 500.$ nm, with waveguide width $w = 542$ nm, waveguide thickness $h = 220.$ nm, slot width $s = 40.$ nm, and slot length $l = 408$ nm. While the periodic hole radius is given by $r = 0.378\,a$, the five holes closest to the slot are linearly tapered in both radius and spacing to 65.6% of their nominal value ( Fig. 1 (b)).

*3.2 Fabrication*

The fabrication process is similar to that previously published [30]. We first pattern the negative resist HSQ004 (80-nm thickness) on silicon-on-insulator (SOI) chips with a top-silicon thickness of 220 nm and a 3-μm buried oxide layer using 100-keV e-beam lithography (Vistec EBPG 5200ES). Inductively-coupled-plasma reactive ion etching (Oxford Plasmalab System 100 ICP) with $HBr/O_2$ chemistry [34] is employed to transfer the pattern into the top silicon. For device release, broadband ultraviolet photolithography with the positive resist AZ ECI 3027 is used to create openings around the devices while protecting the rest of the chip. Device release is subsequently accomplished by submerging the chip in a mixture of one part concentrated HF ($\geq$ 48%) and five parts standard 7:1 buffered oxide etchant for 195 seconds to remove approximately 1.1 μm of the sacrificial buried-oxide layer under the devices.

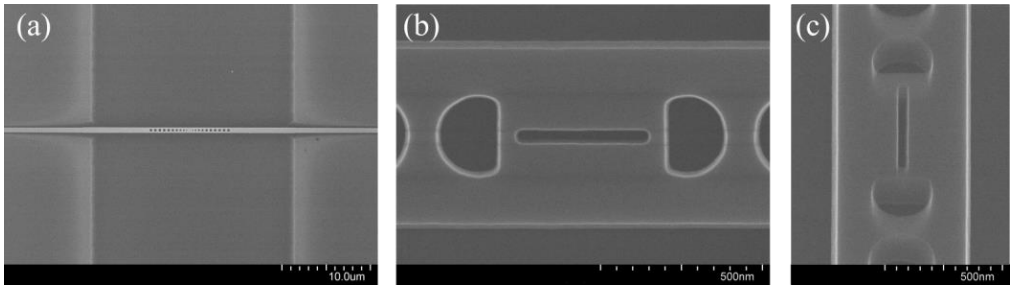

Fig. 2. Scanning electron microscope images of the freestanding photonic crystal nanobeam cavity on a silicon-on-insulator (SOI) wafer with a 40-nm slot. (a) Full device with nine holes on either side of the cavity. The center of the structure, where the cavity is located, is freestanding. The connected waveguides rest on the $SiO_2$ substrate. (b) The central portion of the photonic crystal cavity. (c) View down the slot at 30° tilt. Note the almost perfectly straight inner sidewalls of the slot.

Scanning electron microscope (SEM) images of the resulting structure are shown in Fig. 2. Design dimensions are accurately reproduced in the fabricated devices (typical deviations are $\lesssim$ 10 nm as measured by scanning electron microscopy) with low surface roughness, although there is some deviation from the desired vertical sidewalls, particularly the outer waveguide sidewalls. The most critical dimension in our design however is the slot width. Simulations indicate that a deviation of 5 nm from the target value can reduce the optical quality factor by 80%, which poses a considerable challenge given typical process fluctuations. To meet this stringent requirement, we fabricate structures with a range of nominal slot widths. Slots of length $\lesssim$ 100 nm or width $<$ 30 nm cannot be reliably opened during dry etching, and a 40-nm slot width represents a lower limit for reproducible device fabrication in our process.

The number of holes on each side of the slot is also varied between eight and 11. A higher number of holes decreases the rate of loss of intracavity photons to the attached waveguides, but it also lowers the rate of coupling into the cavity. The chip layout also includes input and output ridge waveguides with a length of several hundred microns connected to each device, at the ends of which focusing grating couplers are located to enable individual testing of the devices as previously described [30]. Here, however, we integrate the photonic crystal cavities into one arm of a Mach-Zehnder interferometer formed by two directional couplers connected with two waveguides (Fig. 3). The directional coupler on the input side of the photonic crystal cavity is designed to send 90% of the light into the cavity and 10% through the waveguide without a cavity. The directional coupler on the output side is a 50:50 splitter.

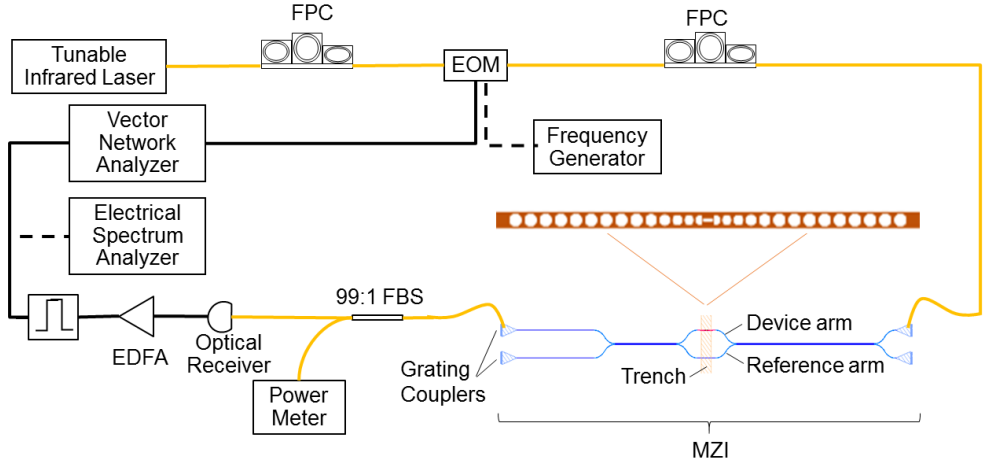

Fig. 3. Schematic of the experimental apparatus for characterization of the photonic crystal nanobeam cavities. See text for definition of abbreviations.

### 3.3 Measurement apparatus

A schematic of the measurement apparatus is shown in Fig. 3. All measurements are performed at atmospheric pressure with the sample chip resting on an aluminum block held at 20.0°C as measured with an integrated thermistor and controlled with a Peltier element. Continuous-wave infrared light (New Focus Velocity 6328 external-cavity tunable diode laser, linewidth (50 ms) < 300 kHz) is directed through a cleaved single-mode fiber into a grating coupler at one input of the Mach-Zehnder interferometer (MZI). A fiber polarization controller (FPC) aligns the polarization of the light with the TE design orientation of the grating couplers. The photonic crystal cavity is located in one arm of the interferometer, permitting homodyne signal detection: The light in the reference arm interferes with the light of the device arm. The resulting signal is collected with a second cleaved single-mode fiber at one output of the MZI.

Depending on the type of measurement made, the light is then routed in various ways. Simple transmission spectra are recorded using a power meter (EXFO IQ 1600). The same power meter is also employed for power monitoring, for which a small part of the transmitted signal is coupled out through a fiber beam splitter (FBS). For measurement of mechanical resonances, the output light is first amplified with an erbium-doped-fiber amplifier (EDFA) (JDS Uniphase MAP EDFA) followed by a narrow bandpass filter (JDS Uniphase MAP Tunable Filter TB3) and then detected with a fast photodiode (JDS Uniphase MAP Receiver RX10) connected to an electrical spectrum analyzer (HP8563A). Calibration-tone measurements are performed by introducing an electro-optical phase modulator (EOM) (Thorlabs 10 GHz LN65S-FC) after the light source. A frequency generator (HP 8341A Synthesized Sweeper 0.01 − 20 GHz) provides the radio-frequency (RF) input signal for the modulator. Finally, for the observation of optomechanically induced absorption, the RF input signal is instead generated with a vector network analyzer (HP8510B), which also analyzes the signal from the fast photodiode.

## 4. Results

### 4.1 Thermomechanical radio-frequency spectra

Determination of the mechanical resonances of a device is accomplished by detecting the RF modulation spectrum of light transmitted through the photonic crystal cavity. The Brownian thermal motion of the mechanical modes modulates the intracavity light. The resulting intensity modulation of the output light leads to Lorentzian resonances in the power spectral density $S_V$ of the transduced photocurrent produced by the fast photodiode, as illustrated in Fig. 4. Here transduced photocurrent refers to the voltage measured at the output of the transimpedance amplifier attached to the photodiode. Each resonance peak corresponds to a different mechanical mode. The mode with the highest amplitude at $\Omega_m/2\pi = 2.69$ GHz is assigned to the breathing mode and matches well the simulated frequency of 2.694 GHz. The inset in Fig. 4 shows a magnification of this resonance. The linewidth of $\Gamma_m/2\pi = 6.4$ MHz corresponds to a mechanical quality factor of $Q_m = 417$. Due to heating of the device as a result of two-photon absorption and the concomitant thermo-optical shift to lower cavity frequency, the laser frequency is blue detuned from the cavity resonance during measurement. Optomechanical backaction in this case leads to an amplification of the mechanical mode and therefore to a slight decrease of the mechanical linewidth [8]. The intrinsic mechanical quality factor is presumably smaller than the above value. In addition to the resonances shown in Fig. 4, there are further less pronounced mechanical resonances up to frequencies as high as $\Omega_m/2\pi = 6.04$ GHz.

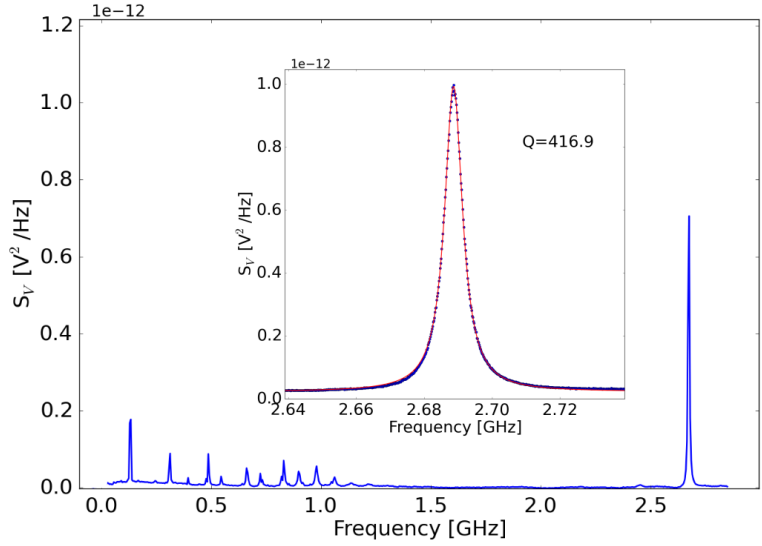

Fig. 4. Power spectral density after the transimpedance amplifier of the fast photodiode. Each resonance peak corresponds to a mechanical mode that couples optomechanically to the light trapped in the cavity. The most prominent peak is the breathing mode at $\Omega_m/2\pi = 2.69$ GHz. The inset shows a magnification of this resonance peak.

*4.2 Optomechanical vacuum coupling rate*

*4.2.1 Calibration tone measurement*

The optomechanical vacuum coupling rate can be determined by taking advantage of the fact that the photonic crystal cavity transduces laser frequency fluctuations and cavity frequency fluctuations in the same way. Following the method described by Gorodetksy et al. [35], we determine $g_0$ by comparing the power spectral density resulting from the thermomechanical cavity frequency fluctuation (Eq. (1)) with that produced by a calibration tone (Eq. (2)).

$$S_\omega(\Omega) \approx 8g_0^2 n_{th} \cdot \frac{\Omega_m^2}{(\Omega^2-\Omega_m^2)^2+\Gamma_m^2\Omega_m^2} \tag{1}$$

$$S_\omega^{cal}(\Omega) = \frac{1}{2}\Omega_{cal}^2\beta^2\delta(\Omega-\Omega_{cal}) \tag{2}$$

$n_{th} \simeq k_B T/\hbar\Omega_m$ denotes the phonon number of the mechanical resonance at frequency $\Omega_m$ for an ambient temperature $T$, where $k_B$ is Boltzmann's constant. We generate the calibration tone by phase modulating the infrared laser with a known modulation depth $\beta$ (see Appendix II) at frequency $\Omega_{cal}$. Both signals lead to a change in photocurrent at the photodiode which gives the power spectral density $S_V(\Omega) = |G_{V\omega}(\Omega)|^2 \cdot S_\omega(\Omega)$, where $G_{V\omega}(\Omega)$ is a frequency dependent transduction factor. Comparing the areas beneath the calibration tone $\langle V^2 \rangle_{cal} = \frac{1}{2}\Omega_{cal}^2 \beta^2 |G_{V\omega}(\Omega_{cal})|^2$, and the thermal noise peak, $\langle V^2 \rangle_m = 2g_0^2 n_{th} |G_{V\omega}(\Omega_{cal})|^2$, we can determine $g_0$ from Eq. (3).

$$g_0 = \frac{\beta \Omega_{cal}}{2} \sqrt{\frac{1}{n_{th}} \frac{\langle V^2 \rangle_m}{\langle V^2 \rangle_{cal}}} \left| \frac{G_{V\omega}(\Omega_{cal})}{G_{V\omega}(\Omega_m)} \right| \qquad (3)$$

Choosing $\Omega_m$ and $\Omega_{cal}$ to differ by only a few megahertz, we can assume the transduction factor to be nearly the same and therefore $\left|\frac{G_{V\omega}(\Omega_{cal})}{G_{V\omega}(\Omega_m)}\right| \approx 1$. An example of such a calibration tone measurement is shown in Fig. 5. Based on the values $\Omega_m/2\pi = 2.69$ GHz and $\beta = 3.4 \times 10^{-3} \pm 0.5 \cdot 10^{-3}$, we infer an optomechanical coupling rate of $g_0/2\pi = 310 \pm 47$ kHz for a device with nine holes on each side of the cavity.

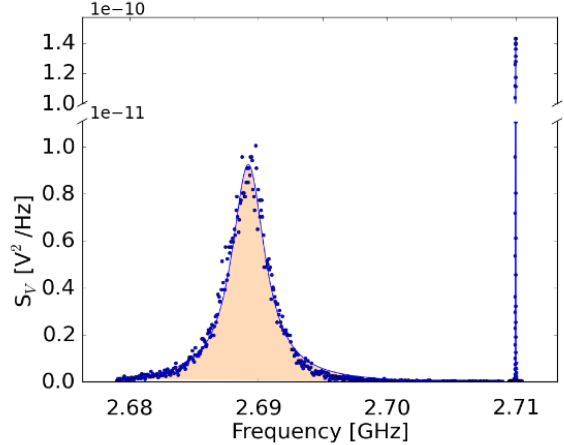

Fig. 5. Determination of $g_0$ by comparing the thermomechanical cavity frequency fluctuation with a calibration tone, which is generated by phase modulating the input laser field. The left peak is the Lorentzian mechanical resonance resulting from the thermal motion of the optical cavity. The right peak is the Gaussian calibration tone.

### 4.2.2 Optomechanically induced absorption

Optomechanically induced absorption (OMIA) is indicative of a coherent interaction of the mechanical motion and the laser field entering the cavity. To observe this behavior, the laser field at frequency $\omega_c$, which we call the control field, is blue detuned from the cavity resonance $\omega_o$. An additional probe field, much weaker than the control field, is created by phase modulating the laser. This leads to two sidebands, of which only the lower frequency one enters the optical cavity to any significant extent, as the upper one is far from resonance. The control field is modulated by the thermal motion of the cavity itself. The lower sideband created by the mechanical motion at $\omega_c - \Omega_m$, interferes constructively with the intracavity probe field, if the probe-control detuning $\Delta_{pc} = \omega_c - \omega_p$ coincides with the mechanical resonance frequency $\Omega_m$. This leads to an enhanced transmission signal.

The expected signal measured by the vector network analyzer (VNA) can be calculated quantitatively [12]. In a frame rotating at $\omega_c$, the transmission coefficient $t_p(\Delta_{pc})$ of the probe signal as a function of the probe-control detuning $\Delta_{pc}$ is given by Eq. (4)

$$t_p(\Delta_{pc}) = \frac{\frac{\kappa_e}{2}}{i(\Delta_{oc}+\Delta_{pc}) + \frac{\kappa}{2} + \frac{G^2}{i(\Omega_m - \Delta_{pc}) - \frac{\Gamma_m}{2}}} \quad (4)$$

where $\kappa_e$ is the decay rate through the Bragg mirrors of the cavity, $\Delta_{oc} = \omega_o - \omega_c$ is the cavity-control detuning and $G = g_0\sqrt{N}$ is the pump-enhanced optomechanical coupling rate for $N$ intracavity photons. The transmitted electric field contains contributions at the carrier and sideband frequencies:

$$E_{out} = e^{i\omega_c t}\left\{t_p(0) + t_p(\Delta_{pc})\frac{\beta}{2}e^{i\Delta_{pc}t} + t_p(-\Delta_{pc})\frac{\beta}{2}e^{-i\Delta_{pc}t}\right\} \quad (5)$$

The signal detected by the photodiode is proportional to $|E_{out}|^2$. The component oscillating with frequency $\Delta_{pc}$ is $I_{\Delta_{pc}} \propto \{|t_p(-\Delta_{pc})|\cos(\Delta_{pc}t + \varphi_-) + |t_p(\Delta_{pc})|\cos(\Delta_{pc}t + \varphi_+)\}$, where $t$ is time, $\beta$ is the modulation depth for the probe sidebands and $\varphi_-$ and $\varphi_+$ are the phase shift experienced by the lower and upper sideband, respectively. The VNA analyzes the in-phase (Eq. (6)) and quadrature component (Eq. (7)) of this signal:

$$I = |t_p(-\Delta_{pc})|\cos(\varphi_-) + |t_p(\Delta_{pc})|\cos(\varphi_+) \quad (6)$$

$$Q = |t_p(-\Delta_{pc})|\sin(\varphi_-) - |t_p(\Delta_{pc})|\sin(\varphi_+). \quad (7)$$

The $|S_{21}|$ parameter displayed by the VNA is proportional to $\sqrt{I^2 + Q^2}$. Ideally $\varphi_+ - \varphi_- = \pi$ for perfect phase modulation. In practice, the sidebands acquire a relative phase during propagation [36]. So $\varphi_+$ is substituted by $\varphi_- + \theta$, and both $\varphi_-$ and $\theta$ are used as fit parameters.

By fitting the model described above with the measured data from the VNA, various parameters can be extracted, e.g. the pump-enhanced optomechanical coupling rate $G = g_0\sqrt{N}$. To determine the vacuum coupling rate $g_0$, the intracavity photon number $N$ is required. This can be determined from the power leaving the cavity $P_{out} = 2\pi N\hbar\omega_o^2/2Q_e$ [37]. The external optical quality factor $Q_e = \omega_o/\kappa_e$ is obtained from the transmitted signal power at a known detuning $\Delta_{oc}$ (Eq. 8) [38].

$$T(\Delta_{oc}) = \frac{1/4Q_e^2}{(\Delta_{oc}/\omega_o)^2 + 1/4Q_o^2} \quad (8)$$

As $P_{out}$ cannot be measured immediately after the photonic crystal cavity but only after several optical elements with their associated losses, e.g. the grating couplers, the measured number for $N$ has considerable uncertainty, which in turn leads to a quite large uncertainty in the inferred value of $g_0$. The losses at the grating couplers and Mach-Zehnder interferometers are estimated by characterizing device structures without a photonic crystal cavity.

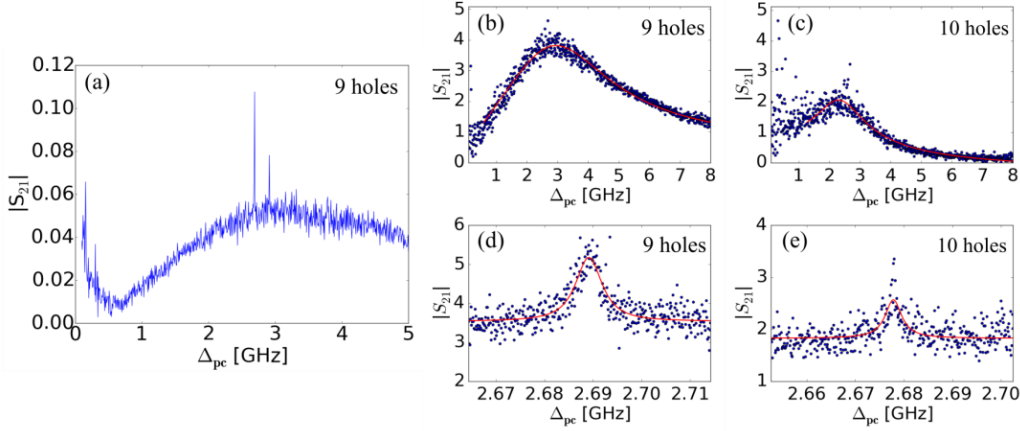

Fig. 6. Measured transmission versus two-photon detuning $\Delta_{pc}$. (a) The two spikes are OMIA signals with the broad background given by the cavity transmission profile. (b) and (c) Same measurement as in (a) with increased range for $\Delta_{pc}$. The measurements were taken for devices with different numbers of holes on each side of the cavity as indicated. The OMIA spikes are not resolved because of the limited number of sampling points. (d) and (e) Measurement at higher resolution of the transmitted signal over the window of induced absorption.

Typical OMIA signals are presented in Fig. 6. Transmission versus two-photon detuning $\Delta_{pc}$ is shown in Fig. 6(a) for a device with nine holes on each side of the cavity. The background is given by the transmission profile of the optical resonance. The two spikes are produced by OMIA at two different mechanical resonances, one at $\Omega_m/2\pi = 2.69$ GHz and the other at $\Omega_m/2\pi = 2.92$ GHz. Data for the same measurement but with a larger range for the two-photon detuning are displayed in Fig. 6(b). Due to the limited number of points recordable by the VNA, the OMIA peaks are not resolved in this measurement, but the full cavity transmission profile is visible. The transmission profile is narrower and more symmetric for a device with ten holes on each side of the cavity (Fig. 6(c)). Indeed, the asymmetry of the signal is expected to decrease with decreasing $\kappa/\Omega_m$ [36]. Figs. 6(d) and 6(e) show a zoom-in of the transmitted signal on the OMIA peak. To reduce the number of fitting parameters necessary to describe the complete transmission profile, we first fit the cavity resonance to the $|S_{21}|$ data shown in Fig. 6(b) and (c) [39] with $G$ set to zero in the expression for $t_p(\Delta_{pc})$ (Eq. (6)). The values extracted for $\kappa$ and $\Delta_{oc}$ are then used to fit the OMIA peak to the data of Figs. 6(d) and 6(e). For the measurements presented here, $\Delta_{oc} \cong -\Omega_m$, which is consistent with the symmetric shape of the OMIA peak [9]. We obtain the following results for the nine-hole device: $Q_o = 5 \times 10^4$, $\kappa/2\pi = 3.9$ GHz and $g_0/2\pi = 600 \pm 300$ kHz. For the ten-hole device, $Q_o = 1.2 \times 10^5$, $\kappa/2\pi = 1.6$ GHz and $g_0/2\pi = 900 \pm 600$ kHz.

## 5. Discussion and conclusion

We have determined the optomechanical vacuum coupling rate $g_0$ for the slotted photonic crystal cavities to be greater than 300 kHz using two independent methods. The OMIA and calibration tone techniques yield results, which are consistent within the range of uncertainty of the measurements and generally in agreement with the simulated value of $g_0/2\pi = 342$ kHz. The calibration tone result however is more reliable for the experiments presented here because the measurement does not depend on knowing the intracavity photon number $N$. In the case of the ten-hole device, the cavity decay rate $\kappa/2\pi = 1.6$ GHz is smaller than the mechanical resonance frequency $\Omega_m/2\pi = 2.69$ GHz, fulfilling the requirement for being in the resolved-sideband regime and opening the door for experiments where primarily one thermomechanical sideband interacts with the intracavity photons, for example, efficient cooling and heating of the mechanical mode.

Our photonic crystals are the first intentionally designed for a high moving-boundary ($g_{mb}$) and small photoelastic ($g_{pe}$) contribution to the overall optomechanical coupling, with a ratio of $g_{mb}:g_{pe} = 6:1$. The approach of collocating the regions of maximum strain with nodes of the electric field leads to a substantial reduction of the photoelastic effect. The value of $g_0$ obtained here is approximately one third of that for the design published by Chan et al. [6], which is dominated by the photoelastic effect. Our simulations indicate, however, that $g_0$ can be increased above 1 MHz simply by going to a slot width of 30 nm, which should be achievable by optimizing the fabrication process. In any case, our design represents a noteworthy alternative, as it can be used with materials that do not have a significant photoelastic effect and the devices should exhibit minimal inherent dispersion, especially when the optical resonance frequency is far from the electronic bandgap.

The small mode volume combined with the electric field maximum being located outside of the silicon offers an interesting opportunity for experiments involving coupling to an additional mechanical resonator. For example, one could investigate the mechanical motion of carbon nanotubes, nanowires or even atoms placed in or near the slot, where they would interact strongly with the optical mode of the cavity and, in turn, couple to mechanical modes of the silicon.

Finally, other known techniques for improving the mechanical behavior of the devices could be exploited. For example, operation at cryogenic temperatures should increase the intrinsic mechanical quality factor [40]. One could also engineer a mechanical bandgap around the frequency of the breathing mode [6]. Both would reduce $\Gamma_m$ and increase the single photon cooperativity $C = g_0^2/(\kappa\Gamma_m)$. Furthermore, thermal effects, which are particularly pronounced in these slotted photonic crystal devices [30], may be reduced by performing pulse-pump measurements [12] or measuring in a buffer gas cryostat [41].

**Appendix I − Calculation of the optomechanical vacuum coupling rate**

From first order perturbation theory, the optomechanical vacuum coupling rate $g_0$ can be described by Eq. (9) [20],

$$g_0 = -\frac{\omega_o}{2}\frac{\int E\frac{\partial\varepsilon}{\partial x}E^* dV}{\int \boldsymbol{E}\cdot\boldsymbol{D}dV}, \tag{9}$$

where $\boldsymbol{E}$ is the electric field, $\boldsymbol{D}$ is the electric displacement field and $\varepsilon$ the material permittivity. In the case of the moving boundary contribution to the coupling, $g_{0,mb}$, this expression becomes

$$g_{0,mb} = -\frac{\omega_o}{2}\frac{\int(\boldsymbol{q}\cdot\hat{\boldsymbol{n}})(\Delta\varepsilon E_\parallel^2 - \Delta\varepsilon^{-1}\boldsymbol{D}_\perp^2)dS}{\int \boldsymbol{E}\cdot\boldsymbol{D}dV} \tag{10}$$

For the cavities described here, $\Delta\varepsilon = \varepsilon_1 - \varepsilon_2$ is the difference between the dielectric constants of silicon and air, and $\Delta\varepsilon^{-1} = \varepsilon_1^{-1} - \varepsilon_2^{-1}$. $\boldsymbol{q}$ is the normed displacement and $\hat{\boldsymbol{n}}$ the vector normal to the surface $S$ between silicon and air. The photoelastic effect in an anisotropic medium is described by $\frac{\partial\varepsilon}{\partial x} = -\varepsilon_0 n^4 p_{ijkl}S_{kl}$, in which case the photoelastic contribution, $g_{0,pe}$, to the optomechanical coupling is

$$g_{0,pe} = -\frac{\omega_o}{2}\frac{\int E(\varepsilon_0 n^4 p_{ijkl}S_{kl})E^* dV}{\int \boldsymbol{E}\cdot\boldsymbol{D}dV} \tag{11}$$

with $\boldsymbol{p}$ being the photoelastic tensor and $\boldsymbol{S}$ the strain tensor [41].

The relative magnitude of the photoelastic and moving boundary contributions depends on the frequency of the light. The photoelastic effect is significantly more dispersive, particularly near the electronic bandgap. This is a result of the photoelastic coefficients $p_{ijkl}$ increasing towards the bandgap [43] and the fact that the effect scales with the refractive index as $n^4$, which increases dramatically near the electronic bandgap. In contrast, the moving boundary effect scales roughly as $n^2$.

**Appendix II − Calibration of the phase modulator**

For the calibration-tone measurement of the optomechanical coupling rate, the modulation depth $\beta = V_p/V_\pi \cdot \pi$ must be known, which requires precise determination of the half-wave voltage $V_\pi$ of the electro-optic phase modulator. $V_p$ is the amplitude of the RF signal applied at the RF input of the EOM. Measurement of $V_\pi$ is accomplished by setting the RF drive signal to 2.9 GHz and varying $V_p$. Light at 1550 nm transmitted through the modulator is passed through a scanning Fabry-Pérot-interferometer (Thorlabs SA210-12B), where the detected signal is resolved into carrier and sidebands. The carrier amplitude as a function of RF power can be described as $a \left(J_0(\pi V_p/V_\pi)\right)^2$ and the first sideband amplitude as $a \left(J_1(\pi V_p/V_\pi)\right)^2$, where $J_0$ and $J_1$ are Bessel functions of the first kind and $a$ is a scaling factor. Fitting the amplitudes of the carrier and sidebands gives $V_\pi = 5.26 \pm 0.07$ V. This value is in agreement with the specifications of the vendor.

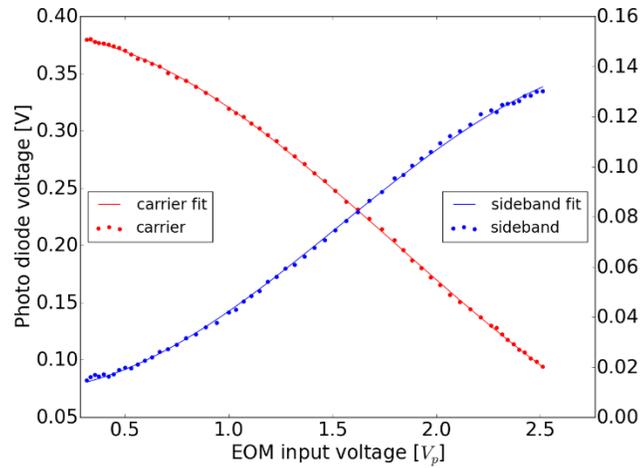

Fig. 7. Measured amplitude of the carrier (red dots) and first sideband (blue dots) as a function of the input voltage at the EOM. Lines are fits of Bessel functions of the first kind to the data.

**Acknowledgments**

We gratefully acknowledge Antonis Olziersky for his expert contributions to the e-beam lithography. The authors also thank Tobias Kippenberg, Dalziel J. Wilson and Nicolas Piro Mastracchio for valuable discussions. K. Schneider was supported by the Marie Curie Initial Training Network "Cavity Quantum Optomechanics" (cQOM) of the European Commission's Seventh Framework Programme (Project ID 290161).